\pdfoutput=1 

\documentclass[a4paper,11pt]{article}

\usepackage{contribution}

\newcommand{\weblink}[2][]{%
    \ifthenelse{\equal{#1}{}}%
    {\textnormal{\url{#2}}}%
    {\textnormal{\href{#2}{#1}}}%
}

\def\beq{\begin{equation}}
\def\eeq#1{\label{#1}\end{equation}}
\def\eeqn{\end{equation}}

\def\beqa{\begin{eqnarray}}
\def\eeqa#1{\label{#1}\end{eqnarray}}
\def\eeqan{\end{eqnarray}}

\let\bar=\overbar

\def\Dslash{\not{\hbox{\kern-4pt $D$}}}
\def\dslash{\not{\hbox{\kern-2pt $\del$}}}

\def\msb{{\bar{\ssstyle M \kern -1pt S}}}

\newcommand{\contribution}[7][]{%
  \clearpage
  \thispagestyle{plain}
  \ifthenelse{\equal{#1}{}}
  {\hypersetup{pdftitle={#2}}}
  {\hypersetup{pdftitle={#1}}}
  \hypersetup{pdfauthor={{#3} {#4}}}
  {\centering\normalfont\LARGE\bfseries\sffamily #2 \par\nobreak}
  \lhead{}
  \chead{%
    \textit{\footnotesize XIV International Conference on Hadron Spectroscopy
      (\weblink[\textit{hadron2011}]{http://www.hadron2011.de}), 13-17 June 2011, Munich, Germany}%
  }
  \rhead{}
  \bigskip
  \begin{center}
    {#3} {#4}\ifthenelse{\equal{#6}{}}{}{\footnote{\weblink[#6]{mailto:#6}}}
    \ifthenelse{\equal{#7}{}}{}{#7} \\
    \textit{#5}
  \end{center}
  \bigskip
}

\renewcommand{\abstract}[1]{%
  \begin{center}
    \begin{minipage}{0.85\textwidth}
      \begin{footnotesize}
        #1
      \end{footnotesize}
    \end{minipage}
  \end{center}
  \bigskip
}

\begin{document}


{  

\contribution[Charmonium(-like) States at Belle]  
{Results on Charmonium and Charmonium-like States at the Belle Experiment}  
{Jens S\"oren}{Lange}  
{Justus-Liebig-Universit\"at Giessen, II.\ Physikalisches Institut, \\
  D-35392 Giessen, GERMANY}  
{soeren.lange@exp2.physik.uni-giessen.de}  
{, on behalf of the Belle Collaboration}

\abstract{%
New results of the Belle experiment at the KEKB asymmetric $e^+$$e^-$ collider 
are presented, in particular {\it (a)} measurement of the mass and width of the $\eta_c$ and $\eta_c'$ 
in $B$ meson decays,
{\it (b)} measurement of the mass, width and quantum numbers of the X(3872) and {\it (c)} observation
of the $\chi_{c2}$ in $B$ meson decays.
}

\section{$\eta_c$ and $\eta_c'$ in $B$ meson decays}

The $\eta_c$ is the 1$^1S_0$ ground state of charmonium with quantum numbers $J^{PC}$=$0^{-+}$.
The $\eta_c'$ represents the first radial excitation 2$^1S_0$.
As a long-standing puzzle the width of the $\eta_c$ 
has been determined with large discrepancies between experiments
with different production mechanisms: in $J$/$\psi$ and $\psi'$ radiative decays
$\Gamma_{\eta_c}$$\simeq$15~MeV, in $B$ meson decays or
$\gamma$$\gamma$$\rightarrow$$\eta_c$ $\Gamma_{\eta_c}$$\simeq$30~MeV \cite{pdg}.
One possible reason is the fact that in radiative decays the cross section
is varying with the photon energy according to $E_{\gamma}^a$ with an exponent
3$\leq$$a$$\leq$7, and thus leading to a distorted line shape of the observed
$\eta_c$ signal. However, in the case of the latter production mechanisms a Breit-Wigner 
lineshape is considered a valid parametrisation.

In a new analysis of $B^+$$\rightarrow$$K^+$$\eta_c$($\rightarrow$$K_S$$K^{\pm}$$\pi^{\mp}$) \cite{etac_belle_2011}, 
the mass and the width of the $\eta_c$ were determined by a 2-dimensional fit of 
the invariant mass $m$($K_S$$K$$\pi$) vs.\ the angle $\angle$($K_S$$K$).
As the $\eta_c$ is a pseudoscalar meson, the angular distribution should be flat. 
However, $P$-wave and $D$-wave components 
by non-resonant charmless $B$ decays 
turned out to be non-negligible. By adding the angle into the fit, 
interference with the background is taken into account.
The mass was determined as $m$=2985.4$\pm$1.5$_{-2.0}^{+0.2}$~MeV.
The measured width in listed in Tab.~\ref{tetac}, in comparison 
with other recent measurements.

\begin{table}[tb]
\begin{center}
\begin{tabular}{|l|l|l|}  
\hline
$\Gamma_{\eta_c}$ & Production Mechanism & Reference \\
\hline
\hline
35.1$\pm$3.1$_{-1.6}^{+1.0}$~MeV & $B$ decays & \cite{etac_belle_2011} and this paper \\
\hline
30.5$\pm$1.0$\pm$0.9~MeV & $\psi'$$\rightarrow$$\gamma$$\eta_c$ & \cite{etac_bes3} \\
\hline
28.1$\pm$3.2$\pm$2.2~MeV & $\gamma$$\gamma$$\rightarrow$$\eta_c$ & \cite{etac_gg_belle} \\
\hline
31.7$\pm$1.2$\pm$0.8~MeV & $\gamma$$\gamma$$\rightarrow$$\eta_c$ & \cite{etac_gg_babar} \\
\hline
36.3$_{-3.6}^{+3.7}$$\pm$4.4~MeV & $B$ decays & \cite{etac_bdecay_babar} \\
\hline
\end{tabular}
\caption{Width measurements of the $\eta_c$.}
\label{tetac}
\end{center}
\end{table}

The analysis was repeated for the $\eta_c'$. 
The measurement of the width of the $\eta_c'$ is of high importance,  
as due to the vicinity to the $D^0$$\overline{D}^0$ threshold, potential model predictions
are not reliable. In case of the $\eta_c'$ the interference with the non-resonant 
background turned out to even have a higher impact for the fit and thus the 
determination of the width. The result is 
$\Gamma$=6.6$_{-5.1}^{+8.4}$$_{-0.9}^{+2.6}$~MeV
for the fit with interference and 
$\Gamma$=41.1$\pm$12.0$_{-10.9}^{+6.4}$~MeV
for a fit without interference 
(i.e.\ fit of only the invariant mass).
The factor $\simeq$5 narrower width of the $\eta_c'$ compared to the $\eta_c$ can be explained
by the wavefunctions of the states. The hadronic decay of both states proceeds by
two gluons; three gluons are forbidden by parity. As the width scales 
with the wavefunction at the origin, i.e.\  
$\Gamma$($^1S_0$$\rightarrow$$g$$g$)$=$(32$\pi$$\alpha_S^2$/$m_c^2$)$|$$\psi$($r$=0)$|^2$,
and the wavefunction for the $\eta_c'$ has one node (as it is $n$=1 radial excitation), 
the width at the origin must be narrower.
With the new measurement, the error on the previous world average of the width
of the $\eta_c'$ was improved by factor $\simeq$2.
For additional details of the analysis see \cite{etac_belle_2011}.

\section{Mass and Width of X(3872)}

New results for the charmonium-like state X(3872) 
in the decays $B^+$$\rightarrow$$K^+$X(3872) 
and $B^0$$\rightarrow$$K^0$($\rightarrow$$\pi^+$$\pi^-$)X(3872) 
are based upon the complete Belle data set of 711~fb$^{-1}$
collected at the $\Upsilon$(4S) resonance \cite{x3872_belle_2011}. 
For the determination of the mass and the width
of the X(3872) in the decay X(3872)$\rightarrow$$J$/$\psi$$\pi^+$$\pi^-$,
a 3-dimensional fit was performed using the three variables
beam constraint mass $M_{\rm bc}$=$\sqrt{(E_{beam}^{cms})^2-(p_B^{cms})^2}$ 
(with the energy in the center-of-mass system $E_{beam}^{cms}$ and the momentum
of the $B$ meson in the center-of-mass system $p_B^{cms}$), 
the invariant mass $m$($J$/$\psi$$\pi^+$$\pi^-$)
and the energy difference $\Delta$$E$=$E_B^{cms}$$-$$E_{beam}^{cms}$ 
(with the energy of the $B$ meson in the center-of-mass system $E^{cms}_B$).
In a first step, the fit was performed for 
the reference channel $\psi'$$\rightarrow$$J$/$\psi$$\pi^+$$\pi^-$,
and the resolution parameters 
(i.e.\ the widths of a core Gaussian and a tail Gaussian)
were then fixed for the fit of the X(3872).
Fig.~1 shows the data and the fits for the X(3872) 
(blue line: signal, dashed green line: background) 
in the projections of the three variables as defined above.
The yield is 151$\pm$15 events for $B^+$ decays
and 21.0$\pm$5.7 events for $B^0$ decays.

{\it Mass of the X(3872).} 
The mass, as determined by the fit, is listed in Tab.~\ref{tx3872} in comparison 
to other precise measurements.
As the X(3872) does not fit into any potential model prediction,
it was discussed as a possible S-wave $D^{*0}$$\overline{D}^0$
molecular state. In this case, the binding energy $E_b$ would be given 
by the mass difference $m$(X)$-$$m$($D^{*0}$)$-$$m$($D^0$). 
Including the new Belle result, the new world average mass of the X(3872)
is $m$=3871.67$\pm$0.17~MeV.
Using the current sum of the masses $m$($D^0$)+$m$($D^{*0}$)=3871.79$\pm$0.30~MeV \cite{pdg},
a binding energy of \mbox{$E_b$=$-$0.12$\pm$0.35~MeV} can be calculated,
which is surprisingly small.
As $E_b$ is inverse proportional to the squared scattering length $a$,
and the radius can in first order be approximated by $<$$r$$>$=$a$/2 \cite{braaten},
this would indicate a very large radius of the molecular state.

\begin{figure}[htb]
\begin{center}
\includegraphics[width=0.6\textwidth,bb=0 0 1060 504]{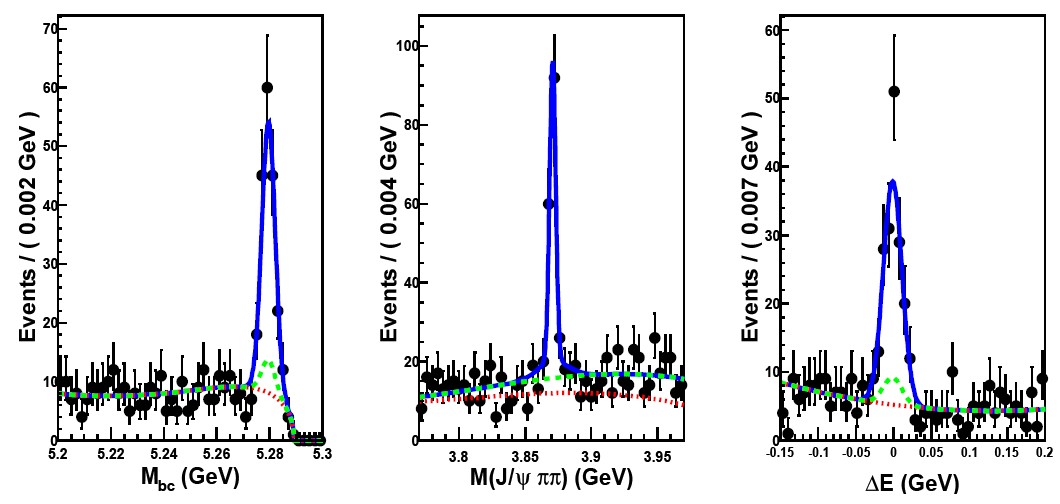}
\caption{
Beam constraint mass $M_{\rm bc}$ (left), 
invariant mass $m$($J$/$\psi$$\pi^+$$\pi^-$) (center)
and $\Delta$$E$ (right) 
for $B^+$$\rightarrow$$K^+$X(3872)($\rightarrow$$J$/$\psi$$\pi^+$$\pi^-$).
}
\label{fx3872}
\end{center}
\end{figure}

\begin{figure}[htb]
\begin{center}
\includegraphics[width=0.43\textwidth,bb=0 0 872 454]{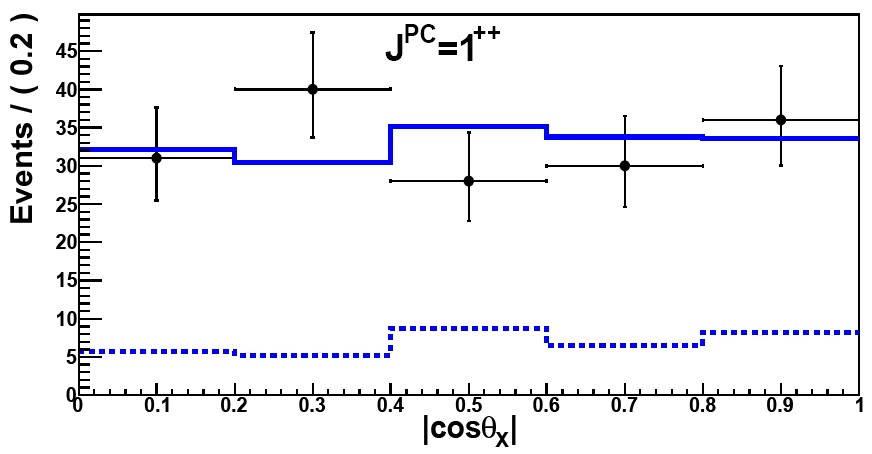}
\includegraphics[width=0.43\textwidth,bb=0 0 875 457]{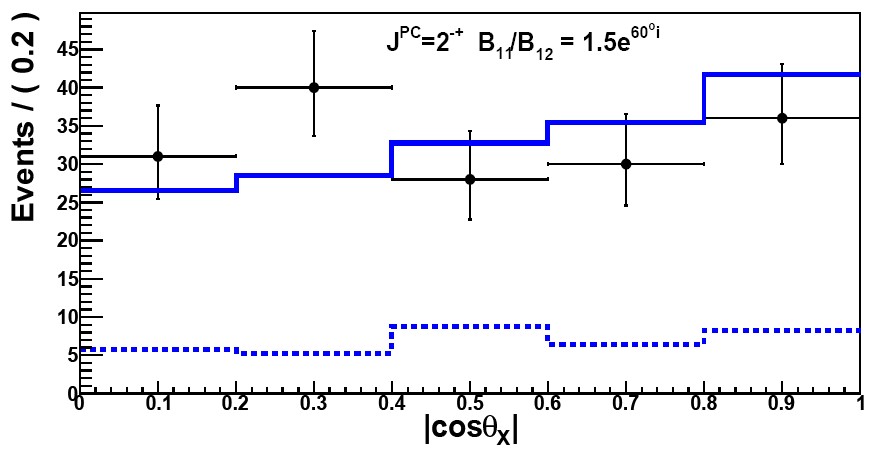}
\caption{
Distribution of $|$cos($\theta_X$)$|$ 
for $B^+$$\rightarrow$$K^+$X(3872)($\rightarrow$$J$/$\psi$$\pi^+$$\pi^-$).
The blue line shows the fit for $J^{PC}$=1$^{++}$ (left) and $J^{PC}$=2$^{-+}$ (bottom).
For details see text. 
}
\label{fx3872_angular}
\end{center}
\end{figure}

\begin{table}[tb]
\begin{center}
\begin{tabular}{|l|l|l|}  
\hline
Experiment & Mass of X(3872) & \\
\hline
\hline
CDF2 & 3871.61$\pm$0.16$\pm$0.19~MeV & \cite{x3872_cdf2} \\
\hline
BaBar ($B^+$) & 3871.4$\pm$0.6$\pm$0.1~MeV & \cite{x3872_babar} \\
\hline
BaBar ($B^0$) & 3868.7$\pm$1.5$\pm$0.4~MeV & \cite{x3872_babar} \\
\hline
D0 & 3871.8$\pm$3.1$\pm$3.0~MeV & \cite{x3872_d0} \\
\hline
Belle & 3871.84$\pm$0.27$\pm$0.19~MeV & \cite{x3872_belle_2011} and this paper \\
\hline
LHCb & 3871.96$\pm$0.46$\pm$0.10~MeV & \cite{x3872_lhcb} \\
\hline
\hline
New World Average & 3871.67$\pm$0.17~MeV & \\
\hline
\end{tabular}
\caption{Mass measurements of the X(3872).}
\label{tx3872}
\end{center}
\end{table}

{\it Width of the X(3872).} 
With the 3-dimensional fit, also a new measurement of the width 
of the X(3872) was performed.
Previously the best upper limit was 
$\Gamma_{X3872}$$<$2.3~MeV (90\% C.L) \cite{x3872_belle_2003}.
The 3-dimensional fits are more sensitive to the natural width
than the resolution provided by the detector $<$$\sigma$$>$$\simeq$4~MeV
because of the constraints which enter by $M_{\rm bc}$ and $\Delta$$E$.
As in case of the mass measurement above, the method of determining 
the width was validated using the $\psi'$ as reference, providing 
a result of $\Gamma_{\psi'}^{measured}$=0.52$\pm$0.11~MeV.
As the world average is $\Gamma_{\psi'}^{PDG}$=0.304$\pm$0.009~MeV,
this indicates a bias in our measurement of $\Delta$$\Gamma$=$+$0.23$\pm$0.11~MeV.
The procedure for the determination of the upper limit is as follows:
for a given fixed width $\Gamma$ the number of signal events
and the number of peaking background events is kept floating in the 3-dim fit, and the 
likelihood is calculated. Then the 90\% likelihood interval is 
determined by finding $w_{90\%}$ for an integral $\int_0^{w_{90\%}}$$\Gamma$$d\Gamma$=0.9.
This procedure gives $w_{90\%}$=0.95~MeV, for which the bias has to be added,
so that $\Gamma_{X(3872)}$$<$1.2~MeV at 90\% C.L. is the final result. 
This upper limit is a factor of $\simeq$2 narrower than the previous upper limit.

{\it Quantum numbers of the X(3872).} 
If the X(3872) is a conventional charmonium state, there are two likely assignments.
On the one hand there is the $\chi_{c1}'$, a $^3P_1$ state with $J^{PC}$=1$^{++}$.
The predicted mass by potential models is $m$=3953~MeV, thus $\simeq$70~MeV
higher than the observed $X$(3872) mass.
This would be a $n$=2 radial excitation, and the quantum numbers are favoured 
by angular analyses \cite{x3872_angular_cdf2} \cite{x3872_angular_belle}.
On the other hand there is the $\eta_{c2}$, a $^1D_2$ state with $J^{PC}$=2$^{-+}$.
The predicted mass by potential models is $m$=3837~MeV, thus $\simeq$35~MeV
lower than the observed $X$(3872) mass.
This would be a $n$=1 state, and the quantum numbers are favoured 
by the 3$\pi$ mass distribution in the decay X(3872)$\rightarrow$$J$/$\psi$$\omega$ \cite{x3872_babar_2010}.
A new angular analysis was carried out with the new Belle data.
For this purpose, it was assumed that the decay X(3872)$\rightarrow$$J$/$\psi$$\pi^+$$\pi^-$
proceeds via X(3872)$\rightarrow$$J$/$\psi$$\rho$($\rightarrow$$\pi^+$$\pi^-$) in the kinematic
limit, i.e.\ both particles are at rest in the X(3872) rest frame. Due to
$m_{X(3872)}$$\simeq$$m_{\rho}$+$m_{J/\psi}$ this is a valid assumption and
it also implies that any higher partial waves can be neglected. 
For $J^{PC}$=1$^{++}$, there is only one amplitude with $L$=0 and $S$=1, 
where $L$ and $S$ are the total orbital angular momentum between and the total spin 
constructed from the $\rho$ and the $J$/$\psi$. 
For $J^{PC}$=2$^{-+}$, there are two amplitudes with $L$=1 and
$S$=1 or $S$=2. 
These two amplitudes can be mixed with a mixing parameter $\alpha$, which is a complex number.
The angular reference frame follows the definition of Rosner \cite{angular_rosner}.
The angle $\theta_X$ is chosen as the angle between the $J$/$\psi$ and
the kaon direction in the $X$(3872) rest frame.
The angular distributions for $\theta_X$ for the different quantum numbers is given by:

\begin{eqnarray}
J^{PC}=1^{++}, & & \frac{d\Gamma}{d cos\theta_X} \propto const.\ \\\nonumber
J^{PC}=2^{-+}, & \alpha=0, & \frac{d\Gamma}{d cos\theta_X} \propto \sin^2 \theta_X \\\nonumber
J^{PC}=2^{-+}, & \alpha=1, & \frac{d\Gamma}{d cos\theta_X} \propto 1 + 3 \cos^2 \theta_X \\\nonumber
\end{eqnarray}

Two additional angles are defined as follows:
the $xy$-plane is spanned by the kaon direction and the $\pi^+$ and $\pi^-$ (back-to-back) directions
in the X(3872) rest frame.
The $x$-axis is chosen to be along the kaon direction.
The $z$-axis is constructed perpendicular to the $xy$-plane.
The angle $\chi$ is chosen between the $x$-axis and the $\pi^+$ direction.
The angle $\theta_{\mu}$ is chosen between the $\mu^+$ direction 
and the $z$-axis.
A simultaneous fit for all three angles was performed, and the distributions and the fit results
for $\theta_X$ are shown in Fig.~\ref{fx3872_angular}. The $\chi^2$ values are listed in Tab.~\ref{tchi2}.
For the case of $J^{PC}$=2$^{-+}$, 
the values in Tab.~\ref{tchi2} are given for $\alpha$=0.69$\cdot$exp($i$23$^o$),
which was found in a grid search and 
which is the only value which gives a confidence level $>$0.1 for all three angles.
Although at the current level of statistical significance, it cannot be distinguished definitely 
between the two quantum numbers, however $J^{PC}$=1$^{++}$ seems to be slightly preferable
in this analysis. For additional details see \cite{x3872_belle_2011}.

\begin{table}[tb]
\begin{center}
\begin{tabular}{|l|r|r|r|r|}  
\hline
Angle & $\chi^2$/n.d.f. & C.L. & $\chi^2$/n.d.f. & C.L. \\
\hline
\hline
 & \multicolumn{2}{|c|}{$J^{PC}$=$1^{++}$} & \multicolumn{2}{|c|}{$J^{PC}$=$2^{-+}$} \\
\hline
\hline
$\chi$ & 1.76/4 & 0.78 & 4.60/4 & 0.33 \\
\hline
$\theta_{lepton}$ & 0.56/4 & 0.97 & 5.24/4 & 0.26 \\
\hline
$\theta_X$ & 3.82/4 & 0.51 & 4.72/4 & 0.32 \\
\hline
\end{tabular}
\caption{$\chi^2$ values for the fit of the angular distributions. 
See text for the definitions of the angles.}
\label{tchi2}
\end{center}
\end{table}

\section{$\chi_{c2}$ in $B$ Meson decays}

In the decay $B^+$$\rightarrow$$K^+$$\chi_{c1,2}$($\rightarrow$$J$/$\psi$$\gamma$)
for the first time a $\chi_{c2}$ signal could be observed with a statistical significance
of 3.6$\sigma$ (Fig.~\ref{fchic2}). This is the observation of a $J$=2 charmonium state with positive parity
in $B$ meson decays and thus very interesting for two reasons:
on the one hand, due to the $j_q$=1/2 of the two charm quarks forming the charmonium state, and the $J$=0 
in the initial state (i.e.\ $J^P$=$0^-$ for the $B$ meson), $J$=0 and $J$=1 are preferred, and 
$J$=2 is difficult to be generated. One the other hand, 
this decay $0^-$$\rightarrow$$0^-$$2^+$ is, because of the positive parity of the charmonium state, 
forbidden in na\"{\i}ve factorization \cite{bauer_stech_wirbel}. 
This implies that at least one additional gluon is required to connect the charmonium and the $K^+$ sides.
For additional details of the analysis see \cite{chic2_belle_2011}.

\begin{figure}[htb]
\begin{center}
\includegraphics[width=0.6\textwidth,bb=0 0 1060 396]{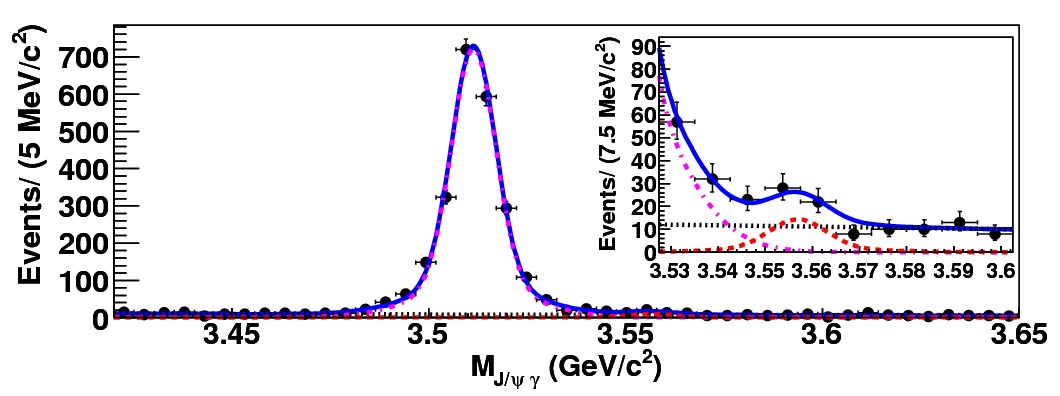}
\caption{
Invariant mass $m$($J$/$\psi$$\gamma$) for the decay 
$B^+$$\rightarrow$$K^+$$\chi_{c1,2}$($\rightarrow$$J$/$\psi$$\gamma$). 
The zoomed region shows the $\chi_{c2}$ signal. See text for details.
}
\label{fchic2}
\end{center}
\end{figure}

\section{Summary}

This paper covered three different topics. 
At first, the width of the $\eta_c'$ was determined with a factor $\simeq$2 smaller error compared
to the previous world average. Interference with non-resonant background 
turned out to be important and were taken into account. 
At second, new results on the X(3872) employed multi-dimensional fits, increasing by constraints 
the resolution to beyond the detector resolution. The new world average mass of the 
X(3872) is only 120$\pm$350~keV below the $D^{*0}$$\overline{D}^0$ threshold.
At third, the production of a $J^P$=$2^+$ charmonium state was observed in $B$ meson decays.



}  

\end{document}